\documentclass[twocolumn,aps,prd,epsf]{revtex4}
\usepackage{epsfig}
\usepackage{amsmath}

\begin{document}

\title{Microscopic 8-quark study of the antikaon nucleon nucleon systems}
\author{P. Bicudo
}
\affiliation{CFTP, Dep. F\'{\i}sica, Instituto Superior T\'ecnico,
Av. Rovisco Pais, 1049-001 Lisboa, Portugal}
\begin{abstract}
We study the possibility to bind eight quarks in a molecular hadronic
system composed of two nucleons and an antikaon, with the quantum numbers of a 
hexaquark flavour, in particular with strangeness -1, isospin 1/2, parity -, 
baryonic number 2 and two possible spins, 0 or 1.
These exotic hadrons are motivated by the deuteron, a proton-neutron boundstate,
and by the model of the $\Lambda(1405)$ as an antikaon proton boundstate.
We discuss the possible production of this hadron in the experiments
which are presently investigating hot topics like the $\Theta^+$ pentaquark or 
the $K^-$ deeply bound in nuclei. 
The $K^- \bullet N$ interactions and the coupling to other channels are computed 
microscopically from a confining and chiral invariant quark model
resulting in local plus separable Gaussian potentials.
The $N \bullet N$ interactions used here are the state of the art 
Nijmegen potentials. The binding energy and the decay rate of the
$K^- \bullet N$ and $K^- \bullet N \bullet N$ systems are computed 
with configuration space variational methods. 
The only systems that bind
with our microscopic interaction are the $K^- \bullet N$ in the $I=0$ channel
and the $K^- \bullet N \bullet N$ in the $S=0$ channel.
\end{abstract}

\maketitle

\section{Introduction}

Here we study the binding energy of a possible molecular antikaon, nucleon and
nucleon three-body system. Importantly, from a hadronic perspective,  
a bound $K^- \bullet N \bullet N$ can be at most reduced to a hexaquark,
a exotic system.
In a nuclear perspective, the deep binding of antikaons
in nuclei has been predicted and are also actively searched. A bound 
$K^- \bullet N \bullet N$ would constitute a simple antikaon-nucleus system,
the ${}^2_{K^-} H$ .

\begin{figure}
\begin{picture}(230,135)(0,0)
\put(0,-5){\includegraphics[width=0.95\columnwidth]{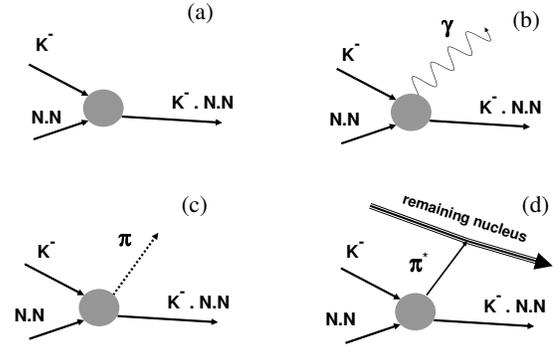}}
\end{picture}
\caption{\label{production} Different $K^- \bullet N \bullet N$ production mechanisms .}
\end{figure}

Moreover, the $K^- \bullet N \bullet N$ can be formed with antikaon $(K^-)$ 
deuteron $(p \bullet n)$ scattering. Other exotic tetraquarks, pentaquarks or
hexaquarks are also 
very plausible, but they are all harder to produce experimentally because they
would need at least strangeness and charm. The several experiments dedicated to 
pentaquark searches (where not only the Kaon, but also the antikaon may interact
with nuclei), or to antikaon-nuclear binding at RCNP, JLab, KEK, DAFNE
and at many other laboratories, are already able to search for 
the proposed $K^- \bullet N \bullet N$. In particular evidence for 
$K^- \bullet N \bullet N$ has already been found by the FINUDA collaboration
at DAFNE
\cite{Agnello:2005qj,Agnello:2006gx}. 
In Fig. \ref{production} different possible production mechanisms are depicted. They
are similar to the $\Lambda(1405)$ production mechanisms, except that the 
$K^-$ scatters on a deuterium nucleus, not on a hydrogen nucleus. The process
in Fig. \ref{production} (a) is only possible if the width of the 
$K^- \bullet N \bullet N$ is of the order of its binding energy. The process
in Fig. \ref{production} (b) is always possible, but is suppressed by the
electromagnetic coupling. Processes in Fig. \ref{production} (c) , (d) are
dominant. The production process (c) should occur in experiments designed
for the production of the pentaquark $\Theta^+$, where the $K^-$ has sufficient
energy to create a $\pi$. The production process (d) should occur in atomic 
Kaon experiments, where the $K^-$ has a low energy but where a larger
nucleus is used and the remaining nucleus may absorb the virtual pion.

The $K^- \bullet N \bullet N$ binding is suggested by the deuteron
${}^2 H$, a proton-neutron boundstate, and by the model of the $\Lambda(1405)$ 
as an antikaon proton boundstate. 
However the subtleties of these two-body subsystems require a precise 
computation of the binding energy of the three-body system.   
For instance the binding of the deuteron $(p \bullet n)$ 
requires a d-wave component
\cite{Stoks:1994wp}. Moreover it is not clear yet if the $K^-\bullet N$ binds in an 
$I=1$ $\Sigma$ baryon. Notice that in the $I=0$ channel there is strong evidence
that the $\Lambda(1405)$ is constituted of two poles, respectively dominated by the
$K^-\bullet N$ and the $\pi \bullet \Sigma$ channels
\cite{Magas:2005vu,Oller:2000fj,Jido:2002yz,Garcia-Recio:2002td,Jido:2003cb,Garcia-Recio:2003ks,Hyodo:2002pk,Nam:2003ch}.

The $K^- \bullet N \bullet N$ binding is a very recent topic. While Oset, Toki and 
collaborators 
\cite{Oset:2005sn,Magas:2006fn}
offer a different interpretation to the FINUDA peak, very recent theoretical calculations,
fitting the $K^- \bullet N$ scattering data with effective local potentials
\cite{Ivanov:2005ng,Yamazaki:2006yc,Yamazaki:2006fk}
or with separable potentials
\cite{Shevchenko:2006xy}
agree with a relatively deep and wide $K^- \bullet N \bullet N$ resonance.

Here we specialize in the microscopic quark computation of the binding energy and 
width of the $K^- \bullet N \bullet N$ systems. Section II is dedicated
to the $N \bullet N$ and $K \bullet N$ interactions. 
We compute microscopically the $K \bullet N$ interaction, with
a confining and chiral invariant quark model and using the Resonating Group Method.  
In Section III we calibrate our interactions with the experimental $K^+ \bullet N$ data, 
and we study the binding and decay of $K^- \bullet N$ and $K^- \bullet N \bullet N$
systems. Finally we conclude.

\section{From quarks to the $K \bullet N$ }

There are several very precise models of the $N \bullet N$ interaction
\cite{Stoks:1994wp},
compatible with the experimental $N \bullet N$ phase shifts.
All contain a long-range attractive potential, a medium-range attractive potential, 
an a hard core or short range potential. A picture consistent with QCD emerges
if the long range attraction is due to the Yukawa one-pion-exchange, the medium
range is due to two-pion or one-sigma exchange, while the short
range repulsion is dominated by the quark interactions. Different models exist
because the $^3S_1 \leftrightarrow ^3D_1$ coupling is partly equivalent to the 
attractive part of the $^3S_1$ potential. Moreover the repulsive hard core potential
essentially pushes the wavefunction outside the repulsive core region, and this can be
accomplished for different heights of the repulsive core. Also, separable or local
potentials can be used.

However the $K^- \bullet N$ interaction has not yet been modelled with the 
same detail of the $N \bullet N$ interaction
\cite{Akaishi:2002bg}. 
Notice that not only the Kaons are unstable 
particles, with less experimental results, but also there are large inelastic 
coupled channel effects in the $K^- \bullet N$ scattering. Moreover,
there is strong evidence that the $\Lambda(1405)$ does not have a single
pole with width $\Gamma$ of $40 MeV$, but two narrower poles, one closer to the 
$K^- \bullet N$ threshold and another dominated by the $\pi \bullet \Sigma$ channel.

\par
This leads us to compute the $K^- \bullet N$ interaction
microscopically at the quark level. 
Here we assume a standard Quark Model (QM) Hamiltonian, 
\begin{equation}
H= \sum_i T_i + \sum_{i<j\, , \, \bar i< \bar j} V_{ij} +\sum_{i \, , \, \bar j} A_{i \bar j} \ ,
\label{Hamiltonian}
\end{equation}
where each quark or antiquark has a kinetic energy $T_i$ with a
constituent quark mass, and the colour dependent two-body
interaction $V_{ij}$ includes the standard QM confining term and a
hyperfine term,
\begin{equation}
V_{ij}= \frac{-3}{16} \vec \lambda_i  \cdot  \vec \lambda_j
\left[V_{conf}(r) + V_{hyp} (r) { \vec S_i } \cdot { \vec S_j }
\right] \ . 
\label{potential}
\end{equation}
For the purpose of this paper the details of potential
(\ref{potential}) are unimportant, we only need to estimate its
matrix elements. The hadron spectrum is compatible with,
\begin{equation}
\langle V_{hyp} \rangle \simeq \frac{4}{3} \left( M_\Delta-M_N
\right) \label{hyperfine}
\end{equation}
Moreover we include in the Hamiltonian (\ref{Hamiltonian}) a
quark-antiquark annihilation potential $A_{i \bar j}$. 
Notice that the quark-antiquark annihilation is constrained 
when the quark model produces spontaneous chiral symmetry breaking
\cite{Bicudo_thesis,Bicudo_scapuz,Llanes-Estrada_thesis,Bicudo_PCAC,weall_pipi,Bicudo_piN}.
In the $\pi$ Salpeter equation, the annihilation potential $A$ 
cancels most of the kinetic energy and
confining potential $2T+V$, 
\begin{equation}
\langle A \rangle_{S=0} \simeq 
\langle 2T+V \rangle_{S=0}
\simeq \langle V_{hyp} \rangle \ ,
\label{sum rules}
\end{equation}
leading to a massless pion in the chiral limit.
We stress that the QM of eq. (\ref{Hamiltonian}) not only
reproduces the meson and baryon spectra as quark and antiquark
bound-states,
but it also complies with the PCAC theorems
\cite{Bicudo_PCAC,weall_pipi,Bicudo_piN,Delbourgo,Llanes-Estrada_l1l2}.

%
%
\begin{figure}[t]
\epsfig{file=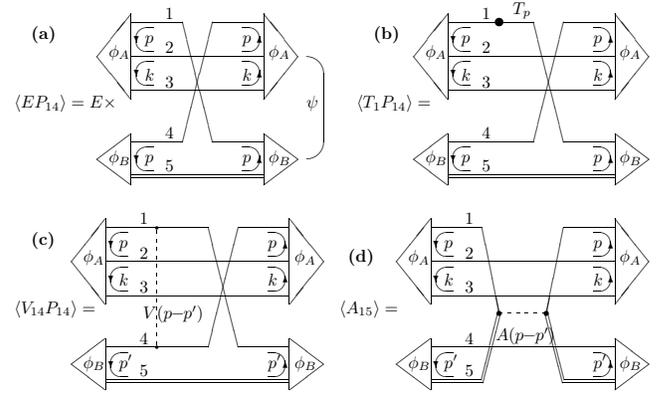,width=8.5cm}
\caption{Examples of our RGM overlaps for the baryon-meson effective 
interaction: (a) norm exchange overlap; (b) kinetic exchange overlap; 
(c) interaction exchange overlap;
(d) annihilation overlap.}
\label{RGM overlaps}
\end{figure}

\par
For the pentaquark system, the Resonating Group Method(RGM)
\cite{Wheeler}
is convenient to arrange the wave functions of quarks and
antiquarks in antisymmetrized overlaps of simple colour singlet
quark clusters, the baryons and mesons.  
The effective potential of the 
meson-baryon system is computed with the overlap of the
inter-cluster microscopic potentials,
\begin{eqnarray}
V_{\text{bar } A \atop \text{mes } B}&=& \langle \phi_B \, \phi_A |
-( V_{14}+V_{1 \bar 5}+2V_{24}+2V_{2 \bar 5} )3 P_{14}
\nonumber \\
&& +3A_{1 \bar 5} | \phi_A \phi_B \rangle /
\langle \phi_B \, \phi_A | 1- 3 P_{14} | \phi_A \phi_B \rangle
\ ,
\label{eq:overlapkernel}
\end{eqnarray}
where $P_{ij}$ stands for the exchange of particle $i$ with
particle $j$, see Fig. \ref{RGM overlaps}. This results in 
an algebraic colour $\times$ spin $\times$ flavour factor and 
a spatial overlap. A convenient basis for 
the spatial wave-functions of the meson $A$ and
the baryon $B$ is the harmonic oscillator basis,
\begin{equation}
\phi_{000}^\alpha(\rho) =  {\cal N_\alpha}^{-1}
\exp\left({ - {{\rho}^2 \over 2 \alpha ^2}}\right) \ , \ \
{\cal N_\alpha} = \left( \sqrt{\pi} \alpha \right)^{3 \over 2} 
\ .
\label{basis}
\end{equation}

In what concerns the quark exchange diagrams, 
the spin independent part of the quark-quark(antiquark) potential 
in eq. (\ref{potential})
essentially vanishes because the clusters are colour singlets. 
The only potential which may contribute is the hyperfine potential. 
The quark Pauli exchange also leads to a positive, repulsive
potential.

Importantly, each annihilation diagram can be related to an exchange 
diagram with the crossing symmetry of one of incoming baryon with the outgoing
baryon. This is evident in Fig. \ref{RGM overlaps} (c) and 
Fig. \ref{RGM overlaps} (d). The corresponding diagrams are opposite
because the annihilation diagram does not have the Pauli exchange minus 
sign of the exchange diagram. 
Another important difference occurs, the exchange diagrams produce
non-local simple separable potentials, while the annihilation diagrams
produce simple local Gaussian potentials. 
Nevertheless all the different diagrams produce hard core potentials
proportional to $\langle V_{hyp} \rangle$,
repulsive in exchange diagrams and attractive in annihilation
diagrams.  Notice that attraction may only occur in non-exotic channels. 

We summarize
\cite{Bicudo:1987tz,Bicudo:1995kq,Bicudo:2004dx,Bicudo_last,Bicudo:2004cm}
the effective potentials computed for the relevant channels,
\begin{eqnarray}
V_{K \bullet N}&=& {c_K}^2
\langle V_{hyp} \rangle  
{\textstyle { 23 \over 32}\left(  1+{ 20 \over 23}  \vec 
\tau_K \cdot \vec
\tau_N \right)\ }
| \phi_{000}^\alpha \rangle \langle \phi_{000}^\alpha | \ ,
\nonumber \\
V_{\bar K \bullet N}({\mathbf r})&=& - {c_K}^2
\langle V_{hyp} \rangle  
{ \textstyle  2 \sqrt 2  
\left( 1-{ 4 \over 3}  \vec 
\tau_K \cdot \vec
\tau_N \right) }
\ e^{-{r^2 / \alpha^2}} \ ,
\nonumber \\
V_{\bar K \bullet N \leftrightarrow \pi \bullet \Lambda}&=& c_\pi c_K
\langle V_{hyp} \rangle  
{ \textstyle { 9 \over 32}\left(  1 +{ 4 \over 3}  \vec 
\tau_K \cdot \vec
\tau_N \right) }
\ | \phi_{000}^\alpha \rangle \langle \phi_{000}^\alpha | \ ,
\nonumber \\
V_{\bar K \bullet N \leftrightarrow \pi \bullet \Sigma}&=& c_\pi c_K
\langle V_{hyp} \rangle  
{ \textstyle {-5 \over 8} \left(  { 1+ \sqrt 6 \over 4} + { -3+ \sqrt 6 \over 3}  \vec 
\tau_K \cdot \vec
\tau_N \right) }\ 
\nonumber \\ && | \phi_{000}^\alpha \rangle \langle \phi_{000}^\alpha | \ ,
\label{K-N}
\end{eqnarray}
where $\vec \tau$ are $1/2$ of the Pauli isospin matrices for the 
$I=0$ and $I=1$ cases, and $c_\pi= \sqrt{E_\pi} f_\pi (\sqrt{2\pi} \alpha)^{3/2}/ \sqrt{3}$ 
is a PCAC factor, and $\mathbf r$ is the relative coordinate.


\par
\begin{figure}[t]
\begin{picture}(100,170)(0,0)
\put(-69.75,1.75){\epsfig{file=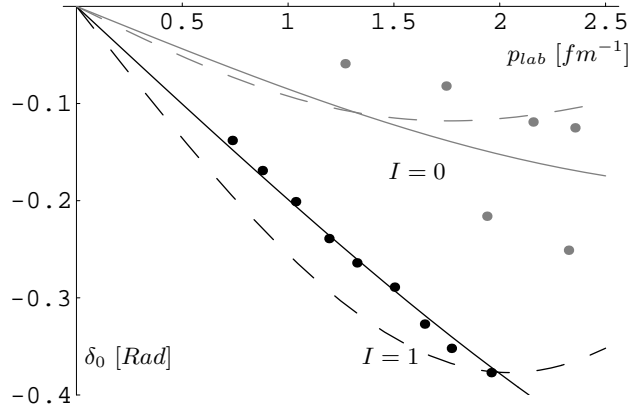,width=8.3cm}}
\put(120,135){$ p_{lab} \ [fm^{-1}] $}
\put(-40,20){$\delta_0  \ [Rad] $} 
\put(75,90){$ I=0$}
\put(65,20){$ I=1$} 
\end{picture}
\caption{
$I=0$ (grey) and $I=1$ (black) s-wave $K \bullet N$ phase shifts as a function 
of the Kaon momentum in the laboratory frame. 
The dots are experimental  
\cite{Hyslop:cs,Barnes,Arndt:2003fj} 
and the solid lines represent this model  
\cite{Bicudo:1987tz,Bicudo:1995kq,Bicudo:2004dx,Bicudo_last,Bicudo:2004cm}
with the hadronic size parameter $\alpha=0.36$ fm and  
${c_k}^2 V_{hyp}= 436 / \sqrt 8=250$ MeV . 
A much larger $\alpha=0.66$ fm, even with a small ${c_k}^2 V_{hyp}=45$ MeV
potential strength, would produce a worse fit (dashed line).
} 
\label{knexp}
\end{figure}

We calibrate our parameters in the two-body $K \bullet N$ channels,
where the diagonalization of the finite difference hamiltonian is straightforward.
From baryon spectroscopy we get 
$\langle V_{hyp} \rangle= 390 MeV$. 
Since $c_K$ is a PCAC factor that suppresses the norm of 
quasi-Golstone-bosons, consistent with the Adler zero in the chiral limit,
we get the plausible parameter interval  
$c_k^2 \langle V_{hyp} \rangle \in [ 200,300]$ MeV.
From the phase shifts for the first channel of eq. (\ref{K-N}), 
the repulsive and well understood  $S=1$ $K \bullet N$ channel
\cite{Hyslop:cs,Barnes,Arndt:2003fj}
depicted in Fig. \ref{knexp}, 
we determine the size parameter $\alpha \in [0.34, 0.39]$ fm.
This is necessarily smaller, by a factor of 2, than the hadronic charge radius 
which is enhanced by the vector meson dominance.

\section{Results and conclusion}

\par
To study the binding and decay  of the $K^- \bullet N$ two-body systems
and of the $K^- \bullet N \bullet N$ three-body systems, we diagonalize the 
hamiltonian in configuration space. Importantly, the decay widths to the channels 
$\pi \bullet \Sigma$ and $\pi \bullet \Sigma \bullet N$ are accounted with 
the substitution method, resulting in an effective two-body $K^- \bullet N$ separable 
potential 
$- |\phi_{000}\rangle  \langle \phi_{000}| G_0(E)|\phi_{000}\rangle \langle \phi_{000}| $.
Because the energy $E$ of our plausible 
resonances is above the pionic channels, the Green function matrix elements are complex. 
This produces the decay width contribution
$- i \Gamma / 2$ to the eigenvalues of the Schr\"odinger two-body and three-body
Hamiltonians.

%
%

First we study the $\Lambda$  channel with the $I=0$ two-body $K^- \bullet N$ 
system. Using the PCAC ratio 
$c_\pi / c_K=\sqrt{ E_\pi / M_K} \simeq 0.56$, 
and solving the $K^- \bullet N$ Schr\"odinger equation in this
parameter range, including the coupling to the $\pi \bullet \Sigma$ 
effective potential, we get a resonance in the $\Lambda(1405)$ region,
with binding energy  
$M-m_K-m_N \in [-23.8,-1.7]$ MeV, and width $\Gamma \in [3.5,7.0]$ MeV. 
This is distant from the single Breit-Wigner model for the
$\Lambda(1405)$ resonance, with a width of $\Gamma\simeq 40$ MeV 
\cite{RPP}
and a binding energy of $-30$ MeV. 
We also consider an increase of the chiral coupling up to $c_\pi / c_K=1$, 
to simulate the possible effect of other decay channels.  
Maintaining the remaining parameter ranges, the resonance 
approaches the $\Lambda(1405)$ region, with binding energy  
$M-m_K-m_N\in [-42.2, -7.7]$ MeV, and width $\Gamma \in [11.2,23.0]$ MeV. 
This is reasonably close to the $\Lambda(1405)$, possibly also close to its double pole 
models, including a slightly narrower Breit-Wigner pole closely below the 
$K^- \bullet N$ threshold (here we don't address the second pole, since for 
simplicity we did not include the $\pi \bullet \Sigma$ interaction, less 
relevant for the $K^- \bullet N \bullet N$ system). 

It is then instructive to find how much we can decrease the $I=0$ attraction, 
and still bind the $K^- \bullet N$ system.
Excluding the coupled channels, the $V_{K^- \bullet N }$ attraction is marginally 
sufficient to bind the $I=0$ system, up to a 1 MeV binding energy. The coupled channel 
effects further bind the system and, even if the overall $I=0$ attraction is reduced 
by a factor $\in[0.77,0.89]$, binding occurs. 

In the $\Sigma$ channel with the $I=1$ two-body $K^- \bullet N$ system, according 
to eq. (\ref{K-N}), the $V_{K^- \bullet N }$ attraction is decreased
by a factor of $ 1 \over 3$, when compared with the $I=0$ channel. Clearly, this
small potential is far from being able to bind the $K^- \bullet N$ onto a $\Sigma$.

\par
We finally study the $K^- \bullet N \bullet N$ three-body system,
using the Nijmegen REID93 potential
\cite{Stoks:1994wp},
together with our  $K^- \bullet N$ interaction and our coupling to the 
$\pi \bullet \Sigma$ and $\pi \bullet \Lambda$ channels.
Importantly, our two-body sub-systems only bind for isospin $I=0$.
For the $N \bullet N$ sub-system we can either have isospin $I_{NN} = 0$ and 
spin $S_{NN} = 1$, or $I_{NN} = 1$ and $S_{NN} = 0$. The two different total $I={1 \over 2}$
systems are respectively the $(-K^- pn + K^-np)/\sqrt 2$ with spin 1, 
and the $(-2 \bar{K}_0 nn - K^- pn - K^- np)/\sqrt 6$  with spin 0. 
In the $S = 1$ case, each $K^- \bullet N$ sub-system is $1 \over 4$ in a $\Lambda$-like $I=0$ state
and $3 \over 4$ in a $\Sigma$-like $I=1$ state.
In the $S = 0$ case, each $K^- \bullet N$ sub-system is 
$3 \over 4$ in a $\Lambda$-like $I=0$ state
and $1 \over 4$ in a $\Sigma$-like $I=1$ state. 
For instance, in the extreme case where the 
two nucleons would be superposed, we can use eq. (\ref{K-N}) to estimate the attraction that this 
dinucleon would impose on the antikaon. In the $S=1$ system, the $K^-$ would feel an
attraction identical to the one it feels in the $\Lambda$ system. In the $S=0$ system, 
the $K^-$ would feel a larger attraction by a factor of $5\over 3$.

We now detail the study of the $S=1$ $K^- \bullet N \bullet N$, where the $N \bullet N$
sub-system binds in a deuteron.
It is
convenient to replace the coordinates of the three 
hadrons $N_1$, $N_2$ and ${K^-}_3$ by relative and centre of mass coordinates
\begin{equation}
\left( \begin{array}{c} \boldsymbol{r}_{12} \\ \boldsymbol{r}_{123} \\ \mathbf{R} \end{array} \right)
= 
\left( \begin{array}{cccc} 
 1 &
 - 1 &
 0
\\
 { 1 \over 2 } &
 { 1 \over 2 }  &
 -1 \\
 { 1 \over 2 + \gamma} &
 { 1 \over 2 + \gamma} &
 { \gamma \over 2 + \gamma} 
\end{array} \right)
\left( \begin{array}{c} \mathbf{r_N}_1 \\ \mathbf{r_N}_2 \\ \mathbf{r_K}_3 \end{array} \right)
\end{equation}
where $\gamma= m_K/m_N$.
The coordinate $\mathbf{R}$ is eliminated in the centre of mass frame,
and we rewrite the hamiltonian 
in terms of scalar products of $ \mathbf{r}_{12}$ , $\mathbf{r}_{123}$ 
and their momenta $\mathbf{p}_{12}$ and $\mathbf{p}_{123}$. 
The total kinetic energy,
\begin{equation}
T={1\over 2 } {2\over  m_N} {\mathbf{p}_{12}}^2
+ {1\over 2 } { 2 + \gamma  \over  2  \gamma  \, m_N }{\mathbf{p}_{123}}^2 \ ,
\end{equation}
is diagonal in $\mathbf{p}_{12}$ and $\mathbf{p}_{123}$. The 
$V_{N_1N_2}$ potential only depends on 
$\mathbf{r}_{12}$, including the tensor interaction. The sum of the other potentials
$V_{N_1{K_3}} + V_{N_2{K_3}}$ is a function of $r_{12}$, $r_{123}$ and of the square of
$\omega = \widehat r_{12} \cdot \widehat r_{123}$.
To span the possible quantum numbers of $K^- \bullet N \bullet N$ system, we apply the total 
potential energy to a groundstate product of s-wave wavefunctions of $\mathbf{r}_{12}$ and 
of $ \mathbf{r}_{123} $. In what concerns the angular quantum numbers, the tensor part of the
$N \bullet N$ interaction $V_{N_1N_2}$ will couple the s-wave in $ \mathbf{r}_{12}$ to 
a d-wave. The other two potentials $V_{N_1{K_3}} + V_{N_2{K_3}}$ couple the groundstate s-wave 
product to wave-functions with any even power of $\omega$. We address the binding of the three-body
$K^- \bullet N \bullet N$ system with two different applications of the variational method. In 
the first method, we use a basis of Laguerre Polynomials in the radial variables $r_{12}$ and
$r_{123}$,
\begin{equation}
\nu_{nl}^\alpha(r)= \sqrt{ {2^{3+2\, l} \, n! \over  (n + 2+2\, l)! } }\, {r^{1+l} \over \alpha^{3/2+ l} }
    \, e^{-\displaystyle r / \displaystyle \alpha } \,  L_n^{2+2\, l}\left(2{r \over \alpha}\right) \ ,
\end{equation}
adequate to study weakly bound systems, together with a basis of Legendre polynomials  $P_{l}(\omega)$ 
in the angular variable $w$. We use up to 10 excitations in each Laguerre basis and 5 excitations
in the Legendre basis, sufficient to get for instance the deuteron binding.
In the second method we use a finite difference method, with 40 excitations 
in each of the radial variables, and 3 in the angular variable. 
The binding and decay width of the $K^- \bullet N \bullet N$ are determined from
the lowest eigenvalue of the hamiltonian, both in the polynomial basis and
in the finite difference basis.

It turns out that in the $K^- \bullet N \bullet N$,  
$I=1/2, \ S=1$ channel there is no binding. The
groundstate has binding in the $r_{12}$ coordinate, but no binding in the
$r_{123}$ coordinate. In particular, the $r_{12}$ part of 
the wavefunction is localized and reproduces the deuteron wavefunction, while the $r_{123}$ 
part is extended over the whole size of the large box where we quantize the wave-function. 
In the limit where the size of the box is infinite, we get a bound deuteron $p \bullet n$ 
and a free $K^-$.  

%
%
\begin{figure}[t]
\begin{picture}(230,140)(0,0)
\put(0,0){
\put(-10,-10){
\includegraphics[width=0.47\textwidth]{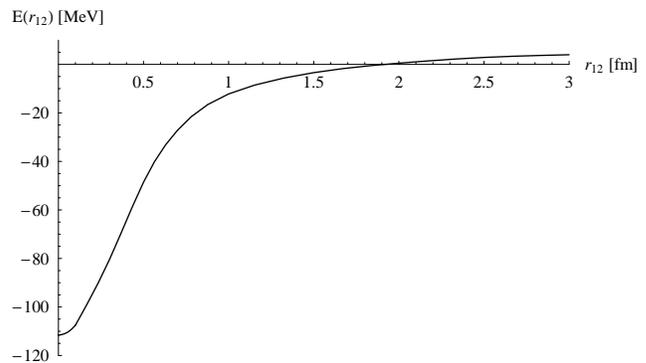}
}
}
\end{picture}
\caption{Kaon energy, or real part of the lowest s-wave eigenvalue of 
$ T_{123} +T_{w} +V_{N_1{K_3}} +V_{N_2{K_3}} $, 
as a function of the nucleon-nucleon distance $r_{12}=|r_{N_1}-r_{N_2}|$. This result is obtained
in a large sphere with a $r_{123}$ radius of 7.5 fm. In the infinite sphere radius limit 
the energy essentially vanishes for $r_{12}>2.0$ fm.}
\label{Kaon attracting nucleons}
\end{figure}

To study in more detail how far we are from binding, we consider the adiabatic 
limit of the $I=1/2, \ S=1$ system, where the positions of the two nucleons are frozen, 
by fixing the $r_{12}$ variable. 
Then the energy of the Kaon $K^-$ can be computed, 
diagonalyzing in detail the terms in the hamiltonian depending on $r_{123}$ and $w$. 
The resulting smaller eigenvalue of $ T_{123} + T_{w}+ V_{N_1 K_3} + V_{N_2 K_3} $ is shown in 
Fig. \ref{Kaon attracting nucleons}, using the parameters in our interval that provide
the strongest possible binding.
It occurs that the antikaon is bound to this frozen deuteron system only if $r_{12} < 2.0 $ fm. 

\begin{figure}[t]
\begin{picture}(230,85)(0,0)
\put(0,0){
\put(-45,5){
\includegraphics[width=0.63\textwidth]{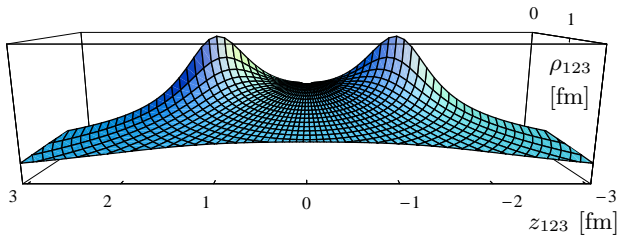}
}
\put(200,0){$z_{123}$ [fm]}
\put(208,60){$\rho_{123}$ }
\put(208,47){[fm]}
}
\end{picture}
\caption{3d perspective (colour online) of the antikaon wavefunction, assuming two 
adiabatically frozen nucleons at the distance of $r_{12}=2.5$ fm, 
obtained with polar coordinates for the $\mathbf r_{123}$, 
with finite differences, and with a radial lattice spacing of $0.1$ fm. 
In this unbound case, the wavefunction is a deformation
of the lowest positive energy harmonic.
}
\label{double maximum}
\end{figure}

Notice that this is much smaller than the deuteron double radius mean square 
$\sqrt{\langle {r_{12}}^2 \rangle }= 3.9$ fm,
which is quite large because at short distances the $N \bullet N$ subsystem
suffers a strong repulsion and because the deuteron is weakly bound. 
Thus is would be very hard to contract the deuteron to a sufficiently small radius
to bind the antikaon. Then, at these large distances, the antikaon essentially feels 
a double-well potential, as shown in Fig. \ref{double maximum}. Importantly,
such a widely separated double-well potential only binds if any of the two wells 
is sufficiently deep to bind the antikaon. It occurs that this already 
happens if the $K^- \bullet N$ attraction 
in the $S=1$ three-body system
is arbitrarily increased by a factor $\in [1.14,1.35]$. Then we get
a vanishing binding energy, and a decay width of $\Gamma \in [8.0,9.8]$ MeV.

This can also be applied to the $S=0$, $K^- \bullet N \bullet N$ three-body system, 
where we have binding because the attraction in the $K^- \bullet N$ sub-systems
is increased by a factor of $5/3$ when compared with the
$S=1$, $K^- \bullet N \bullet N$ three-body system. 
In particular we find
a binding energy $M - m_K-2m_N \in [-53.0,-14.2]$ MeV, 
and a decay width $\Gamma \in [13.6,28.3]$ MeV
to the $\pi \bullet \Sigma \bullet N$
and $\pi \bullet \Lambda \bullet N$ channels.
The complex pole of this resonance is comparable to the one we get for  
the $\Lambda(1405)$.

To conclude we compute, starting at the quark level, the $K^- \bullet N$ 
interactions, constrained by chiral symmetry and by the crossing symmetry to the 
$K^+ \bullet N$ system. We find binding in the $I=0, \ K^- \bullet N$ 
channel, possibly consistent with the double pole model of the $\Lambda(1405)$,
and no binding in the $I=1, \ K^- \bullet N$ two-body $\Sigma$.
We also study the $I=1/2, \ K^- \bullet N \bullet N$ three-body systems.
In the $ S=1, \ K^- \bullet N \bullet N$ system, 
we find no binding, although we almost get binding.
We find a binding in the
$ S=0, K^- \bullet N \bullet N$ system, 
with a complex pole comparable to the one of the $\Lambda(1405)$. 
Essentially, the main difference to the generally larger binding and decay
of very recent theoretical calculations 
\cite{Ivanov:2005ng,Yamazaki:2006yc,Yamazaki:2006fk,Shevchenko:2006xy,Akaishi:2002bg}
can be linked to our shorter range and weaker potential $V_{K^- \, N}$. 

%
%

\vspace{-0.7cm}
 
\acknowledgements

\vspace{-0.3cm}

PB is indebted to Avraham Gal for an algebraic correction.
PB also 
thanks Eulogio Oset and Paola Gianotti for advices and discussions 
and Marco Cardoso and George Rupp for support on the $N \bullet N$ interaction.


\end{document}